# Control of gene expression by modulated self-assembly

Jose M. G. Vilar[1,2] and Leonor Saiz[3,*]

[1]Biophysics Unit (CSIC-UPV/EHU) and Department of Biochemistry and Molecular Biology, University of the Basque Country, P.O. Box 644, 48080 Bilbao, Spain

[2]IKERBASQUE, Basque Foundation for Science, 48011 Bilbao, Spain

[3]Department of Biomedical Engineering, University of California, 451 E. Health Sciences Drive, Davis, CA 95616, USA

## Abstract

**Numerous transcription factors self-assemble into different order oligomeric species in a way that is actively regulated by the cell. Until now, no general functional role has been identified for this widespread process. Here we capture the effects of modulated self-assembly in gene expression with a novel quantitative framework. We show that this mechanism provides precision and flexibility, two seemingly antagonistic properties, to the sensing of diverse cellular signals by systems that share common elements present in transcription factors like p53, NF-κB, STATs, Oct, and RXR. Applied to the nuclear hormone receptor RXR, this framework accurately reproduces a broad range of classical, previously unexplained, sets of gene expression data and corroborates the existence of a precise functional regime with flexible properties that can be controlled both at a genome-wide scale and at the individual promoter level.**

---

[*] Correspondence: lsaiz@ucdavis.edu

# INTRODUCTION

A recurrent theme in gene regulation is the self-assembly of transcription factors (TF) into coexisting populations of dimers, tetramers, and other higher order oligomers that can bind simultaneously single and multiple DNA sites. This behavior has been observed explicitly in the tumor suppressor p53 (1), the nuclear factor κB (NF-κB) (2,3), the signal transducers and activators of transcription (STATs) (4), the octamer-binding proteins (Oct) (5,6), and the retinoid nuclear hormone receptor RXR (7) (Table 1). In these systems, the properties of self-assembly, and the partitioning into low and high order oligomeric species, are strongly regulated and modulated by several types of signals, such as ligand binding (8), protein binding (9,10), acetylation (11), and phosphorylation (6,12). The general implications of this modulation, however, are not clear.

At the level of single DNA sites, it is well established that the effects of TF are finely determined by their concentration and cognate DNA sequences (13). Processes based on interactions with different molecules and post-transcriptional modifications are assumed to affect mainly the DNA binding properties of the TFs or their ability to recruit coregulators. This idea is entrenched in the field of gene regulation and is systematically used as a guiding principle in the ongoing development of molecular therapies against diverse diseases (14). But TFs rarely act through just a single binding site (6,15-21) (Table 1). Modulated self-assembly (MSA) provides a key mechanism for controlling the ability of TFs to bind two or more DNA sites simultaneously.

To determine the common wide-ranging effects of MSA, we have developed a general quantitative framework that accurately links MSA with control of gene expression (Figure 1). It focuses on the general aspects of the core control mechanism shared by the wide variety of regulatory systems where MSA is present, which include TF self-assembly and its modulation, binding of the TF oligomers to DNA, and the resulting transcriptional responses. This quantitative framework allowed us to uncover modulation of the oligomeric states of TFs as a flexible mechanism for precise sensing of molecular signals in the presence of intracellular fluctuations. Precision ensures that the transcriptional response is consistently triggered at a given modulator signal strength irrespective of the TF concentration. Flexibility allows the precise triggering point to be changed, up to several orders of magnitude, both at the individual promoter level by changing its DNA sequence and at a genome-wide scale by changing the molecular self-assembly properties.

This methodology identified a core set of features needed to implement control of transcription by MSA that are present in a wide variety of structurally different systems (Table 1). As an exemplar of these systems, we have considered explicitly the nuclear hormone receptor RXR. In this case, the quantitative framework accurately reproduced, in some instances even without free parameters, a broad range of classical, previously unexplained gene expression experimental data and demonstrated how flexible precise control of gene expression can be achieved directly at the molecular level through modulation of the oligomerization state of transcriptional regulators.

## MATHERIALS AND METHODS

The first step in the signaling cascade orchestrated by MSA is the regulation of the relative abundance of the oligomerization states of the TF (Figure 1). The self-assembly modulator, such as a ligand that binds to a TF or a kinase that phosphorylates the TF, affects the low-order oligomers to promote or prevent their self-assembly. We consider explicitly tetramers, $n_4$, dimers, $n_2$, and non-tetramerizing dimers, $n_2^*$, as relevant high and low order oligomeric species. Other oligomerization pairs, such as octamer-tetramers or dimer-monomers, are mathematically equivalent to tetramer-dimers.

We quantitate the effects of self-assembly modulation through the modulator function $f([s]) = [n_2^*]/[n_2]$, which describes, in terms of concentrations, the partitioning into the tetramerizing and non-tetramerizing dimers by the self-assembly modulator, $s$. This process affects dimer and tetramer concentrations, which are related to each other through $[n_2]^2/[n_4] = K_{td}$, where $K_{td}$ is the tetramer-dimer dissociation constant.

The precise form of the modulator function is given by the specific mode of action of the modulator. An explicit example is $f([s]) = [s]/K_{lig}$ for a ligand $s$ that upon binding to the dimer $n_2$, with dissociation constant $K_{lig}$, renders it unable to tetramerize in the form $n_2^*$. Another relevant, mechanistically different situation corresponds to $f([s]) = k_{dephos}/([s]v_{phos})$ for phosphorylation in the linear regime of the non-tetramerizing, $n_2^*$, into the tetramerizing, $n_2$, dimer species. In this case, $[s]$ is the concentration of active kinases and $v_{phos}$ and $k_{dephos}$ are the phosphorylation and dephosphorylation rate constants, respectively. In general, several mechanisms can be involved at the same time in controlling the oligomerization properties. For instance, the case in which the two previous processes are combined so that the dimer has to be both free of ligand and phosphorylated to be able to tetramerize leads to a two-variable modulator function given by $f([s_l],[s_p]) = [s_l]/K_{lig} + [s_l]k_{dephos}/(K_{lig}[s_p]v_{phos}) + k_{dephos}/([s_p]v_{phos})$, where $[s_l]$ is the ligand concentration and $[s_p]$ is the concentration of active kinases.

Binding of the different TF oligomers to the DNA sites mediates the transcriptional effects of the self-assembly modulator (Figure 1). Typically, tetramers and both types of dimers bind single DNA sites in a very similar way, with free energies $\Delta G_{s1}^o$ and $\Delta G_{s2}^o$, for site 1 and 2, respectively. These quantities are related to the corresponding dissociation constants through $\Delta G_{s1}^o = RT\ln(K_{s1})$ and $\Delta G_{s2}^o = RT\ln(K_{s2})$. Tetramers, in addition, can bind two sites simultaneously because they have two DNA binding domains, one from each of its two constituent dimers, which contribute with $\Delta G_{s1}^o$ and $\Delta G_{s2}^o$ to the free energy. The simultaneous binding of two domains is typically accompanied by conformational changes, e.g. twisting and bending, in both the tetramer and DNA (22,23), which contributes with an additional conformational term, $\Delta G_C^o$, to the free energy. Therefore, the standard free energy of the state with the tetramer bound to two sites is given by $\Delta G_{s1}^o + \Delta G_{s2}^o + \Delta G_C^o$. This conformational contribution has been studied in detail in the case of DNA looping by prokaryotic TFs and is dependent, among others, on the TF and DNA flexibility, the relative position of the DNA binding sites, and the DNA supercoiling state (22,23).

We use statistical thermodynamics to quantitatively describe binding to DNA in terms of free energies and concentrations of the different oligomeric species (24-26). The key quantity is the statistical weight, or Boltzmann factor, defined as $Z_i = [n_4]^{t_i}[n_2]^{d_i}[n_2^*]^{m_i} e^{-\Delta G_i^o/RT}$, which relates the relative probability of the binding state $i$ with its standard free energy $\Delta G_i^o$. The exponents $t_i$, $d_i$, and $m_i$ correspond to the number of tetramers, dimers, and non-tetramerizing dimers in the state $i$, respectively. The factor $RT$ is the gas constant, $R$, times the absolute temperature, $T$. The probability of a given group of binding states $c$, $P_c = \sum_{i \in c} Z_i / \sum_i Z_i$, is obtained by adding the statistical weights of its states and normalizing by the sum for all the possible states.

For a system with two binding sites, there are 17 binding states (Figure 1). These states are those with both sites empty; one occupied by a dimer or a tetramer; two sites occupied by two dimers, by two tetramers, or a dimer and tetramer; and two sites occupied simultaneously by a single tetramer. In the case of states with dimers, one has to take into account that a dimer can either be in the form that allows or prevents tetramerization. In general, each binding state includes a constellation of molecular substates with different DNA conformations. For instance, the state with both sites empty can include a bent DNA conformation, as in the case when the two sites are occupied simultaneously by a single tetramer, but the lack of a tetramer to stabilize the conformation makes this conformation highly unlikely. This type of effects has been described in detail for other TF that bind two DNA sites simultaneously, such as the *lac* repressor (27).

There is also the possibility that oligomerization is so weak in solution that it is only observed on DNA. This effect can be put in quantitative terms with our framework by considering that the state with the tetramer bound simultaneously to two DNA sites (Figure 1) can also be described as two interacting dimers that bind cooperatively to DNA. The statistical weight of this state is given by $Z_2 = [n_4]e^{-(\Delta G_{s1}^o + \Delta G_{s2}^o + \Delta G_C^o)/RT}$ in terms of tetramer concentration and by $Z_2 = ([n_2]^2 / K_{td})e^{-(\Delta G_{s1}^o + \Delta G_{s2}^o + \Delta G_C^o)/RT}$ in terms of dimer concentration, which can be rewritten as $Z_2 = [n_2]^2 e^{-(\Delta G_{s1}^o + \Delta G_{s2}^o + \Delta G_{int})/RT}$ with $\Delta G_{int} = \Delta G_C^o + RT \ln K_{td}$. Thus, a very high dissociation constant that does not lead to significant tetramerization in solution is sufficient to promote tetramerization on DNA when the conformational free energy is sufficiently low. Intuitively, tetramerization is observed on DNA because binding to DNA brings the tetramerization domains close to each other and increases their local concentration.

Two differentiated types of transcriptional responses can be constructed from the binding states of the TF on DNA (Figure 1).

The first type, referred to as response R1, involves a high order oligomer that simultaneously binds two non-adjacent DNA sites. Upon binding, the high-order oligomer loops out the intervening DNA and positions a distal enhancer in the vicinity of the promoter region to control transcription. The probability $P_t$ of the state with the tetramer bound to the two DNA sites simultaneously (Table 2) determines the effective transcription rate through the expression $\bar{\Gamma}_{R1} = \Gamma_{ref}(1-P_t) + \Gamma_t P_t$, which weights the transcription rates that the system has with, $\Gamma_t$, and without, $\Gamma_{ref}$, the distal enhancer close to the promoter.

The second type, denoted here response R2, takes advantage of the differentiated recruitment abilities of different oligomerization states. This mode of regulation applies to a coactivator that is recruited by a low-order oligomer by binding to a molecular surface that is occluded in the high-order oligomer. In this case, the effective transcription rate is given by $\bar{\Gamma}_{R2} = \Gamma_{ref}(1 - P_{do} - P_{od} - P_{dd}) + \Gamma_{do} P_{do} + \Gamma_{od} P_{od} + \Gamma_{dd} P_{dd}$. The subscripts do, od, and dd of the transcription rates $\Gamma$ and probabilities $P$ refer to the group of states with dimers bound to just site 1, to just site 2, and to both sites, respectively (Table 2). $\Gamma_{ref}$ is the transcription rate with no dimers bound, including empty sites and sites occupied by tetramers.

Responses R1 and R2 embrace the prototypical cases mediated by long and short range interactions between regulatory elements. They are controlled by the relative occupancy of DNA binding sites by the different oligomeric species. This mode of functioning differs from other systems with multiple binding sites, like the *lac* operon, which are controlled by the absolute occupancy of their sites by a single oligomeric species (28). For instance, IPTG, an inducer of the *lac* operon, does not affect the oligomerization state of the tetrameric *lac* repressor but prevents each of its two DNA binding domains from significantly binding their cognate sites (28,29).

## RESULTS

To uncover the unique characteristics that emerge from the core structure of MSA in such a general wide variety of structurally different systems (Table 1), we focus on a functional regime that guarantees that there is response to changes in the self-assembly modulator concentration. This regime considers two properties. The first one is that the TF concentration is sufficiently high for it to significantly bind DNA. In mathematical terms, it implies $[n_4]+[n_2]+[n_2^*] \gg e^{\Delta G_{s1}^o/RT}$ and $[n_4]+[n_2]+[n_2^*] \gg e^{\Delta G_{s2}^o/RT}$. The second one is that the tetramer concentration is sufficiently low, $[n_4] \ll [n_2]+[n_2^*]$, so that they do not to take completely over the binding. The reason is that for typical values of $\Delta G_C^o$, tetramers bind more strongly to two DNA sites simultaneously than dimers do to a single DNA site (27,30,31).

The key implication of this regime is that the probabilities of the different groups of binding states simplify in such a way (see Table 2) that the transcriptional responses are governed by the reduced expressions

$$\begin{aligned}\bar{\Gamma}_{R1} &= \Gamma_{ref}(1-P_t)+\Gamma_t P_t \\ \bar{\Gamma}_{R2} &\approx \Gamma_{ref} P_t + \Gamma_{dd}(1-P_t)\end{aligned} \quad (1)$$

with

$$P_t \approx \frac{1}{1+\left(1+f([s])\right)^2 e^{\Delta G_C^o/RT} K_{td}}, \quad (2)$$

which show that responses R1 and R2, despite being mechanistically different, follow the same control logics. In both cases, the two-site binding of the tetramer, quantified by $P_t$, determines the contributions of the reference and activated transcriptional states. The end result is even more remarkable because the particular form of $P_t$ imparts precision and flexibility to the transcriptional responses, two properties that are the cornerstone of natural gene expression systems but that have proved to be highly elusive because of their seemingly antagonistic character (13).

Precision ensures that the transcriptional response is consistently triggered at a given modulator signal strength irrespective of the particular TF concentration, which cancels out in the reduced equations that govern the system behavior. Flexibility, on the other hand, allows the precise triggering point to be altered, up to several orders of magnitude, both at the individual promoter level by changing its organization —$\Delta G_C^o$ depends on the distance between the two DNA binding sites (17,22)— and at a genome-wide scale by changing the molecular self-assembly properties — $f([s])$ and $K_{td}$ affect the regulation of all genes in the same way.

All these results can be observed explicitly in the retinoid X receptor (RXR), an exemplar of the essential regulators that share the central features of MSA (Table 1). RXR controls a large number of genes by binding to DNA as homodimer, homotetramer, or obligatory heterodimerization partner for other nuclear receptors. Nuclear retinoid receptors are highly significant because they mediate the pleiotropic effects of retinoic acid, which include cell proliferation, differentiation, and embryonic development and affect the carcinogenic process in a number of organs (32).

The canonical self-assembly modulator of RXR is the hormone 9-*cis*-retinoic acid (9cRA), a derivative of Vitamin A, which binds each RXR subunit independently of its oligomerization state (33) and prevents dimers with their two subunits occupied from tetramerazing (8). This behavior is consistent with $n_2$ being an apo-dimer and with $n_2^*$ being a holo-dimer, as observed in the respective crystal structures of the dimers with no ligand bound (34) and with two ligands bound (35). The crystal structure of one tetramer with two ligands bound (36) shows that two dimers with just one ligand each can form tetramers with a structure similar to those of two apo-dimers. In addition to 9cRA, there are other ligands of RXR, as for instance, the oleic acid, docosahexaenoic acid, methoprene acid, and phytanic acid (35).

These early steps in sensing 9cRA and other ligand concentrations are taken into account by the explicit form of the modulator function, which we obtain from the mass action law as

$$f([s]) = \frac{[n_2^*]}{[n_2]} = \frac{[s]^2}{K_{lig}^2 + 2K_{lig}[s]}, \tag{3}$$

where $K_{lig}$ and $[s]$ are the ligand-RXR dissociation constant and the ligand concentration, respectively (see Supplementary Methods).

To compare with the experimental data, we normalize the fold induction, a measure of relative changes in transcriptional activity, so that its variation ranges from 0 to 1. This quantity, referred to as normalized fold induction (*NFI*), is defined explicitly as $NFI = (FI-1)/(FI_{max}-1)$, where $FI_{max}$ is maximum value of the fold induction $FI$. In terms of the *NFI*, the results do not depend on parameters related to the baseline and maximum expression levels and it becomes possible to effectively compare experiments on different promoters and cell lines (see Supplementary Methods). The only parameters needed to characterize the shape of the response in the functional regime are $K_{lig}$ and $K_{td}$, which have been measured experimentally, and $\Delta G_C^o$, which can be inferred by adjusting its value to reproduce the experimental data.

This approach accurately describes the experimental observations (16) for the ligand 9cRA and a promoter with two non-adjacent DNA binding sites for RXR and a distal enhancer (Figure 2A). Simultaneous binding of an RXR tetramer to the two sites loops out the intervening DNA and brings the enhancer close to the promoter region (response R1). Increasing the concentration of 9cRA prevents the formation of RXR tetramers and leads to deactivation of transcription.

The very same approach also captures in detail the observed behavior when the two DNA binding sites are next to each other, as in the classic set of experiments that uncovered 9cRA as the cognate ligand of RXR, for different promoters and cell lines (Figure 2B). In these cases, only the dimeric forms of RXR with ligand bound can recruit a coactivator (response R2) and increasing the concentration of 9cRA results in the activation of transcription. The extent of activation is modulated by the RXR AF-1 domain and RXR phosphorylation (37-39).

This framework has the much-sought ability to fully predict, without free parameters, the responses to different ligands from the values of $\Delta G_C^o$ obtained in response to just a single ligand. Applied to the all-trans retinoic acid (atRA), which was tested early on as a potential cognate ligand of RXR (40,41), the approach closely recapitulates its effects on transcription for different cell types and promoters from the values of $\Delta G_C^o$ inferred in the responses to 9cRA (Figure 2C). This ability to fully predict responses without free parameters is especially important because it provides a direct avenue to transfer specific molecular information of the ligand-TF interaction, as described by the measured or computed parameters, across scales up to the transcriptional effects.

The high variability of the transcriptional responses, as observed in Figure 2, has been a long-standing recurrent issue in RXR gene regulation. In particular, the half-maximum response point, characterized by the $EC_{50}$, ranges from just above the RXR-ligand dissociation constant up to values 30-fold higher (Table 3). Our results have identified MSA as a potential mechanism to control the $EC_{50}$ at the single-gene level through the value of $\Delta G_C^o$ (Table 3). This promoter-dependent flexibility indicates that for these systems, the observed variability is not a random aspect of the experimental setup but the result of RXR precisely tailoring the response to each individual gene.

The observed variability can be collapsed in the form of response landscapes (Figure 3), which represent the transcriptional activity as a function of the conformational free energy in addition to just the usual ligand concentration of dose-response curves. The landscapes explicitly show the ability of RXR to shape the molecular response to ligand binding in a promoter-dependent way. The response landscapes show how the $EC_{50}$ increases as the conformational free energy decreases in a way that closely matches the experimental observations (Figure 3).

To investigate the extent to which typical experimental conditions fall within the functional regime (which, as previously described, is characterized by $[n_4]+[n_2]+[n_2^*] \gg e^{\Delta G_{s1}^o/RT}$, $[n_4]+[n_2]+[n_2^*] \gg e^{\Delta G_{s2}^o/RT}$, and $[n_4] \ll [n_2]+[n_2^*]$), we considered the model for RXR in the whole-parameter space. All groups of binding states were considered explicitly without simplifications of the expressions for the corresponding probabilities (Table 2). In addition to the relevant quantities of the functional regime, the whole-parameter space includes the experimentally measured free energies of binding to DNA, the RXR dimer-monomer dissociation constant, and the nuclear RXR concentration. The results (Figure 4) are virtually independent of the precise value of the total nuclear RXR protein concentration over, at least, a 10-fold range and accurately capture the diverse dose-response curves observed in the experiments, in agreement with the results for the functional regime. In all cases, the ranges of concentrations include 550 nM, the estimated RXR nuclear concentration in HL-60 cells (7). Therefore, the ability to elicit flexible and precise responses, as uncovered in the general analysis, is also present when the particularities of RXR-mediated transcriptional responses are taken into account.

## DISCUSSION

Cellular processes rely on intricate molecular mechanisms to function in extraordinarily diverse intra- and extra-cellular environments. Eukaryotic gene expression, in particular, has shown to be exceedingly complex (42-44). Just the core of the transcriptional machinery itself involves a wide variety of components with oscillatory patterns of macromolecular assembly and phosphorylation (45). On top of the constitutive processes, there are many other molecular interactions that provide regulation, enhancing or reducing gene expression and adjusting to changing cellular conditions (46). To understand how these different levels of molecular complexity contribute to the observed behavior, one needs the right approaches (47,48).

The quantitative framework we have developed provides an efficient avenue to connect the molecular properties of MSA with its effects in the control of gene expression. This framework allowed us to uncover unique properties of control of gene expression by MSA that lead to a flexible mechanism for precise sensing of diverse types of self-assembly modulation signals, irrespective of changes in transcription factor concentration. Application of this methodology to the nuclear hormone receptor RXR accurately describes the experimentally observed transcriptional responses for both enhancers (response R1) and coactivators (response R2) from just the molecular properties of the components (Figures 2A and 2B), and successfully predicts the observed behavior without free parameters (Figure 2C). A detailed analysis of the whole-parameter space reveals that regulation by RXR is functioning in a precise regime, with minimal dependence on RXR nuclear concentration (Figure 4), in which the responses are highly diverse as a result of the inherent flexibility that accompanies precision in the control of gene expression by MSA (Figure 3).

The observed TF-concentration insensitivity of control of gene expression by MSA contrasts with the traditional role of RXR as obligatory heterodimerization partner for other nuclear receptors, which relies on the absolute occupancy of the cognate binding sites by the heterodimer. In the case of RXRα:PPARγ regulation of adipogenesis, however, it has been observed that several promoters are controlled rather by the relative occupancy between RXRα:PPARγ heterodimers and other RXRα heterodimers or homo-oligomers (49). Our framework provides a starting point to consider these more complex situations by coupling MSA with hetero-oligomerization and to combine these extensions with recent bioinformatics methods (50,51) to make accurate predictions on gene expression based on the binding profiles observed in the experimental data (49).

The combined presence of flexibility and precision in the control of gene expression by MSA, as explicitly shown for RXR, allows a single TF to simultaneously regulate multiple genes with promoter-tailored dose-response curves that consistently maintain their diverse shapes for a broad range of the TF concentration changes. These features are especially important because essential TFs like p53, NF-κB, STATs, Oct, and RXR, each of which have all the core elements that form the backbone of control of gene expression by MSA, regulate multiple genes that engage in processes as diverse as cancer, inflammation, autoimmune diseases, and cellular differentiation. These results indicate that the prospects for devising more effective molecular therapies for systems controlled by MSA will greatly benefit from shifting potential intervention points from those that affect absolute concentrations and single-site binding to those that can tackle concentration ratios and promoter properties.

## SUPPLEMENTARY DATA

Supplementary Data is available from NAR online.

## FUNDING

This work was supported by the MICINN under grant FIS2009-10352 (J.M.G.V.) and the University of California, Davis (L.S.).

# REFERENCES

1.  Wang, P., Reed, M., Wang, Y., Mayr, G., Stenger, J.E., Anderson, M.E., Schwedes, J.F. and Tegtmeyer, P. (1994) p53 domains: structure, oligomerization, and transformation. *Mol Cell Biol*, **14**, 5182-5191.

2.  Phelps, C.B., Sengchanthalangsy, L.L., Malek, S. and Ghosh, G. (2000) Mechanism of kappa B DNA binding by Rel/NF-kappa B dimers. *J Biol Chem*, **275**, 24392-24399.

3.  Sengchanthalangsy, L.L., Datta, S., Huang, D.B., Anderson, E., Braswell, E.H. and Ghosh, G. (1999) Characterization of the dimer interface of transcription factor NFkappaB p50 homodimer. *J Mol Biol*, **289**, 1029-1040.

4.  Zhang, X. and Darnell, J.E., Jr. (2001) Functional importance of Stat3 tetramerization in activation of the alpha 2-macroglobulin gene. *J Biol Chem*, **276**, 33576-33581.

5.  Tomilin, A., Remenyi, A., Lins, K., Bak, H., Leidel, S., Vriend, G., Wilmanns, M. and Scholer, H.R. (2000) Synergism with the coactivator OBF-1 (OCA-B, BOB-1) is mediated by a specific POU dimer configuration. *Cell*, **103**, 853-864.

6.  Kang, J., Gemberling, M., Nakamura, M., Whitby, F.G., Handa, H., Fairbrother, W.G. and Tantin, D. (2009) A general mechanism for transcription regulation by Oct1 and Oct4 in response to genotoxic and oxidative stress. *Genes Dev*, **23**, 208-222.

7.  Kersten, S., Kelleher, D., Chambon, P., Gronemeyer, H. and Noy, N. (1995) Retinoid X receptor alpha forms tetramers in solution. *Proc Natl Acad Sci U S A*, **92**, 8645-8649.

8.  Chen, Z.P., Iyer, J., Bourguet, W., Held, P., Mioskowski, C., Lebeau, L., Noy, N., Chambon, P. and Gronemeyer, H. (1998) Ligand- and DNA-induced dissociation of RXR tetramers. *J Mol Biol*, **275**, 55-65.

9.  van Dieck, J., Fernandez-Fernandez, M.R., Veprintsev, D.B. and Fersht, A.R. (2009) Modulation of the oligomerization state of p53 by differential binding of proteins of the S100 family to p53 monomers and tetramers. *J Biol Chem*, **284**, 13804-13811.

10. Hanson, S., Kim, E. and Deppert, W. (2005) Redox factor 1 (Ref-1) enhances specific DNA binding of p53 by promoting p53 tetramerization. *Oncogene*, **24**, 1641-1647.

11. Li, A.G., Piluso, L.G., Cai, X., Wei, G., Sellers, W.R. and Liu, X. (2006) Mechanistic insights into maintenance of high p53 acetylation by PTEN. *Mol Cell*, **23**, 575-587.

12. Wenta, N., Strauss, H., Meyer, S. and Vinkemeier, U. (2008) Tyrosine phosphorylation regulates the partitioning of STAT1 between different dimer conformations. *Proc Natl Acad Sci U S A*, **105**, 9238-9243.

13. Ptashne, M. and Gann, A. (1997) Transcriptional activation by recruitment. *Nature*, **386**, 569-577.

14. de Lera, A.R., Bourguet, W., Altucci, L. and Gronemeyer, H. (2007) Design of selective nuclear receptor modulators: RAR and RXR as a case study. *Nat Rev Drug Discov*, **6**, 811-820.

15. Mangelsdorf, D.J., Umesono, K., Kliewer, S.A., Borgmeyer, U., Ong, E.S. and Evans, R.M. (1991) A direct repeat in the cellular retinol-binding protein type II gene confers differential regulation by RXR and RAR. *Cell*, **66**, 555-561.

16. Yasmin, R., Yeung, K.T., Chung, R.H., Gaczynska, M.E., Osmulski, P.A. and Noy, N. (2004) DNA-looping by RXR tetramers permits transcriptional regulation "at a distance". *J Mol Biol*, **343**, 327-338.

17. Jackson, P., Mastrangelo, I., Reed, M., Tegtmeyer, P., Yardley, G. and Barrett, J. (1998) Synergistic transcriptional activation of the MCK promoter by p53: tetramers link separated DNA response elements by DNA looping. *Oncogene*, **16**, 283-292.

18. Stenger, J.E., Tegtmeyer, P., Mayr, G.A., Reed, M., Wang, Y., Wang, P., Hough, P.V. and Mastrangelo, I.A. (1994) p53 oligomerization and DNA looping are linked with transcriptional activation. *EMBO J*, **13**, 6011-6020.

19. Leung, T.H., Hoffmann, A. and Baltimore, D. (2004) One nucleotide in a kappaB site can determine cofactor specificity for NF-kappaB dimers. *Cell*, **118**, 453-464.

20. Vinkemeier, U., Cohen, S.L., Moarefi, I., Chait, B.T., Kuriyan, J. and Darnell, J.E., Jr. (1996) DNA binding of in vitro activated Stat1 alpha, Stat1 beta and truncated Stat1: interaction between NH2-terminal domains stabilizes binding of two dimers to tandem DNA sites. *EMBO J*, **15**, 5616-5626.

21. Xu, X., Sun, Y.L. and Hoey, T. (1996) Cooperative DNA binding and sequence-selective recognition conferred by the STAT amino-terminal domain. *Science*, **273**, 794-797.

22. Saiz, L., Rubi, J.M. and Vilar, J.M.G. (2005) Inferring the in vivo looping properties of DNA. *Proc Natl Acad Sci U S A*, **102**, 17642-17645.

23. Saiz, L. and Vilar, J.M.G. (2006) DNA looping: the consequences and its control. *Curr Opin Struct Biol*, **16**, 344-350.

24. Hill, T.L. (1960) *An Introduction to Statistical Thermodynamics*. Addison-Wesley.

25. Ackers, G.K., Johnson, A.D. and Shea, M.A. (1982) Quantitative model for gene regulation by lambda phage repressor. *Proc Natl Acad Sci U S A*, **79**, 1129-1133.

26. Segal, E. and Widom, J. (2009) From DNA sequence to transcriptional behaviour: a quantitative approach. *Nat Rev Genet*, **10**, 443-456.

27. Vilar, J.M.G. and Saiz, L. (2005) DNA looping in gene regulation: from the assembly of macromolecular complexes to the control of transcriptional noise. *Curr Opin Genet Dev*, **15**, 136-144.

28. Saiz, L. and Vilar, J.M. (2008) Ab initio thermodynamic modeling of distal multisite transcription regulation. *Nucleic Acids Res*, **36**, 726-731.

29. Kuhlman, T., Zhang, Z., Saier, M.H., Jr. and Hwa, T. (2007) Combinatorial transcriptional control of the lactose operon of Escherichia coli. *Proc Natl Acad Sci U S A*, **104**, 6043-6048.

30. Vilar, J.M.G. and Leibler, S. (2003) DNA looping and physical constraints on transcription regulation. *J Mol Biol*, **331**, 981-989.

31. Saiz, L. and Vilar, J.M.G. (2006) Stochastic dynamics of macromolecular-assembly networks. *Molecular Systems Biology*, **2**, 0024.

32. Altucci, L. and Gronemeyer, H. (2001) The promise of retinoids to fight against cancer. *Nat Rev Cancer*, **1**, 181-193.

33. Kersten, S., Dawson, M.I., Lewis, B.A. and Noy, N. (1996) Individual subunits of heterodimers comprised of retinoic acid and retinoid X receptors interact with their ligands independently. *Biochemistry*, **35**, 3816-3824.

34. Bourguet, W., Ruff, M., Chambon, P., Gronemeyer, H. and Moras, D. (1995) Crystal structure of the ligand-binding domain of the human nuclear receptor RXR-alpha. *Nature*, **375**, 377-382.

35. Egea, P.F., Mitschler, A. and Moras, D. (2002) Molecular recognition of agonist ligands by RXRs. *Mol Endocrinol*, **16**, 987-997.

36. Gampe, R.T., Jr., Montana, V.G., Lambert, M.H., Wisely, G.B., Milburn, M.V. and Xu, H.E. (2000) Structural basis for autorepression of retinoid X receptor by tetramer formation and the AF-2 helix. *Genes Dev*, **14**, 2229-2241.

37. Mascrez, B., Mark, M., Krezel, W., Dupe, V., LeMeur, M., Ghyselinck, N.B. and Chambon, P. (2001) Differential contributions of AF-1 and AF-2 activities to the developmental functions of RXR alpha. *Development*, **128**, 2049-2062.

38. Nagpal, S., Friant, S., Nakshatri, H. and Chambon, P. (1993) RARs and RXRs: evidence for two autonomous transactivation functions (AF-1 and AF-2) and heterodimerization in vivo. *EMBO J*, **12**, 2349-2360.

39. Chambon, P. (2005) The nuclear receptor superfamily: a personal retrospect on the first two decades. *Mol Endocrinol*, **19**, 1418-1428.

40. Heyman, R.A., Mangelsdorf, D.J., Dyck, J.A., Stein, R.B., Eichele, G., Evans, R.M. and Thaller, C. (1992) 9-cis retinoic acid is a high affinity ligand for the retinoid X receptor. *Cell*, **68**, 397-406.

41. Levin, A.A., Sturzenbecker, L.J., Kazmer, S., Bosakowski, T., Huselton, C., Allenby, G., Speck, J., Kratzeisen, C., Rosenberger, M., Lovey, A. *et al.* (1992) 9-cis retinoic acid stereoisomer binds and activates the nuclear receptor RXR alpha. *Nature*, **355**, 359-361.

42. Levine, M. and Tjian, R. (2003) Transcription regulation and animal diversity. *Nature*, **424**, 147-151.

43. Simicevic, J. and Deplancke, B. (2010) DNA-centered approaches to characterize regulatory protein-DNA interaction complexes. *Mol Biosyst*, **6**, 462-468.

44. Ptashne, M. and Gann, A. (2002) *Genes & signals*. Cold Spring Harbor Laboratory Press, Cold Spring Harbor, N.Y.

45. Metivier, R., Penot, G., Hubner, M.R., Reid, G., Brand, H., Kos, M. and Gannon, F. (2003) Estrogen receptor-alpha directs ordered, cyclical, and combinatorial recruitment of cofactors on a natural target promoter. *Cell*, **115**, 751-763.

46. Agalioti, T., Lomvardas, S., Parekh, B., Yie, J., Maniatis, T. and Thanos, D. (2000) Ordered recruitment of chromatin modifying and general transcription factors to the IFN-beta promoter. *Cell*, **103**, 667-678.

47. Vilar, J.M.G., Guet, C.C. and Leibler, S. (2003) Modeling network dynamics: the lac operon, a case study. *J Cell Biol*, **161**, 471-476.

48. Lee, E. and Bussemaker, H.J. (2010) Identifying the genetic determinants of transcription factor activity. *Mol Syst Biol*, **6**, 412.

49. Nielsen, R., Pedersen, T.A., Hagenbeek, D., Moulos, P., Siersbaek, R., Megens, E., Denissov, S., Borgesen, M., Francoijs, K.J., Mandrup, S. *et al.* (2008) Genome-wide profiling of PPARgamma:RXR and RNA polymerase II occupancy reveals temporal activation of distinct metabolic pathways and changes in RXR dimer composition during adipogenesis. *Genes Dev*, **22**, 2953-2967.

50. Vilar, J.M.G. (2010) Accurate prediction of gene expression by integration of DNA sequence statistics with detailed modeling of transcription regulation. *Biophys J*, **99**, 2408-2413.

51. Vilar, J.M. and Saiz, L. (2010) CplexA: a Mathematica package to study macromolecular-assembly control of gene expression. *Bioinformatics*, **26**, 2060-2061.

52. Bissonnette, R.P., Brunner, T., Lazarchik, S.B., Yoo, N.J., Boehm, M.F., Green, D.R. and Heyman, R.A. (1995) 9-cis retinoic acid inhibition of activation-induced apoptosis is mediated via regulation of fas ligand and requires retinoic acid receptor and retinoid X receptor activation. *Mol Cell Biol*, **15**, 5576-5585.

53. Vuligonda, V., Lin, Y. and Chandraratna, R.A.S. (1996) Synthesis of highly potent RXR-specific retinoids: The use of a cyclopropyl group as a double bond isostere. *Bioorganic & Medicinal Chemistry Letters*, **6**, 213-218.

54. Widom, R.L., Rhee, M. and Karathanasis, S.K. (1992) Repression by ARP-1 sensitizes apolipoprotein AI gene responsiveness to RXR alpha and retinoic acid. *Mol Cell Biol*, **12**, 3380-3389.

# TABLES

**Table 1.** Modulated self-assembly of transcription factors

| TF[a] | Self-assembly modulation | Oligomerization states[b] | DNA binding |
|---|---|---|---|
| RXR | ligand-binding (8) | monomer, dimer*, tetramer* (7) | 1 site, 4 consecutive half-sites (15), 2 separated sites (16) |
| p53 | protein-binding (9,10), acetylation (11) | monomer, dimer, tetramer*, stacked-tetramers* (1) | 1 site, 2 separated half-sites, 2 separated sites (17,18) |
| NF-κB | protein-mediated | dimer*, tetramer* (2,3) | 2 separated sites (19) |
| STAT | phosphorylation (12) | dimer*, tetramer* (4) | tandem sites (20,21) |
| Oct | phosphorylation (6) | monomer*, dimer* (5), tetramer* (6) | 1 site, 2 separated sites (6) |

[a]For each transcription factor (TF), the table shows the experimentally observed mechanism of the self-assembly modulation process, the oligomerization states involved, and the corresponding arrangement of DNA binding sites at the promoter.
[b]The symbol * indicates the oligomeric species that have been observed to substantially bind DNA.

**Table 2.** Probability, $P_c$, of the different groups of binding states

| $P_c$ | States[a] | Full expression[b] | Simplified expression[c] |
|---|---|---|---|
| $P_t$ | 2 | $\dfrac{e^{-\Delta G_C^o/RT}[n_4]}{e^{-\Delta G_C^o/RT}[n_4]+\left(e^{\Delta G_{s1}^o/RT}+[n_4]+[n_2]+[n_2^*]\right)\left(e^{\Delta G_{s2}^o/RT}+[n_4]+[n_2]+[n_2^*]\right)}$ | $\dfrac{1}{1+\dfrac{\left([n_2]+[n_2^*]\right)^2}{e^{-\Delta G_C^o/RT}[n_4]}}$ |
| $P_{od}$ | 6, 7, 15, 16 | $\dfrac{\left([n_2]+[n_2^*]\right)\left(e^{\Delta G_{s2}^o/RT}+[n_4]\right)}{e^{-\Delta G_C^o/RT}[n_4]+\left(e^{\Delta G_{s1}^o/RT}+[n_4]+[n_2]+[n_2^*]\right)\left(e^{\Delta G_{s2}^o/RT}+[n_4]+[n_2]+[n_2^*]\right)}$ | 0 |
| $P_{do}$ | 9, 10, 12, 13 | $\dfrac{\left([n_2]+[n_2^*]\right)\left(e^{\Delta G_{s1}^o/RT}+[n_4]\right)}{e^{-\Delta G_C^o/RT}[n_4]+\left(e^{\Delta G_{s1}^o/RT}+[n_4]+[n_2]+[n_2^*]\right)\left(e^{\Delta G_{s2}^o/RT}+[n_4]+[n_2]+[n_2^*]\right)}$ | 0 |
| $P_{dd}$ | 8, 11, 14, 17 | $\dfrac{\left([n_2]+[n_2^*]\right)\left([n_2]+[n_2^*]\right)}{e^{-\Delta G_C^o/RT}[n_4]+\left(e^{\Delta G_{s1}^o/RT}+[n_4]+[n_2]+[n_2^*]\right)\left(e^{\Delta G_{s2}^o/RT}+[n_4]+[n_2]+[n_2^*]\right)}$ | $\dfrac{1}{1+\dfrac{e^{-\Delta G_C^o/RT}[n_4]}{\left([n_2]+[n_2^*]\right)^2}}$ |

[a]States involved in the group as described in Figure 1.
[b]The expressions for the probabilities follow from the statistical thermodynamic approach with the free energies of each state as described in Figure 1.
[c]Simplified expressions for the probabilities in the functional regime.

**Table 3.** $EC_{50}$ control by ligand binding strength and conformational free energy

| $EC_{50}$ (nM) | Ligand | $K_{lig}$ (nM) | $\Delta G_C^o$ (kcal/mol) |
|---|---|---|---|
| 287.4 | 9cRA | 8 | 8.03 |
| 77.8 | 9cRA | 8 | 9.47 |
| 18.3 | 9cRA | 8 | 10.76 |
| 14.3 | 9cRA | 8 | 10.92 |
| 3403.2 | atRA | 350 | 9.47 |
| 798.7 | atRA | 350 | 10.76 |
| 626.0 | atRA | 350 | 10.92 |

The $EC_{50}$ is defined as the ligand concentration that gives the half-maximum response.

**FIGURE LEGENDS**

**Figure 1.** Quantitative modeling of control of gene expression by modulated self-assembly. Intracellular signals are processed through *modulated self-assembly* into populations of different oligomeric species that upon *DNA binding* engage in *transcription control*. *Modulated self-assembly*.— The intensity of a self-assembly modulator signal $[s]$, e.g. ligand or active kinase concentration, regulates the formation of high order oligomers by modifying (represented as a yellow spark) the low order oligomers and preventing their self-assembly into the high order species. *DNA binding*.— The oligomeric species bound to DNA (in orange/red) are described by their free energies with the statistical weights ($Z_{state}$) shown for each binding state (expression in black). The parenthesized number, in blue, labels each of the 17 states and the molecular representations illustrate the binding combinations of the transcriptional regulator to the two DNA sites (site 1 and site 2). The top left box summarizes the notation. *Transcriptional control*.— One state (state 2) can trigger response R1, in which an enhancer is positioned in the vicinity of the promoter region, and twelve states (states 6-17) can potentially trigger response R2, in which a coactivator is recruited to the promoter region. Dimers and tetramers have been drawn as compositions of the nuclear hormone receptor RXR structures from the PDB files 1BY4 (DNA binding domains bound to the two half-sites on DNA, or RXR response elements) and 1G1U (ligand binding domains).

**Figure 2.** Prediction of RXR-mediated transcriptional responses to 9cRA and atRA ligands. The results of the model (lines) for the functional regime are compared to the normalized fold induction (*NFI*) from experimental data (symbols) for different promoters and ligands. The model uses the experimental values $K_{lig} = 8$ nM for 9cRA (52) or $K_{lig} = 350$ nM for atRA (53), and $K_{td} = 4.4$ nM (7). The conformational free energy $\Delta G_C^o$ (shown in kcal/mol) is inferred from just the experimental data for 9cRA by minimizing the mean squared error between model and experiments and the resulting

value is used subsequently for responses to atRA. **(A)** Response to 9cRA for a system with two separated DNA binding sites for RXR and a distal enhancer. Experimental gene expression data is taken from Figure 5b of Yasmin et al. (16), which used COS-7 cells transfected with the reporter, consisting of double RXRE and a UAS site 300 bp upstream, in a vector encoding GAL4-VP16. The *NFI* was computed as $NFI_{R1} = P_t$ (see Supplementary Methods) with equations (2) and (3). In this case, the half-maximum response concentration, or $EC_{50}$, is about 35 times higher than the 9cRA-RXR dissociation constant. **(B)** Responses to 9cRA for systems with contiguous DNA binding sites for RXR. The variability of the dose-response curves, including 10-fold changes in the $EC_{50}$ and different slopes, is accurately captured by the model by just adjusting $\Delta G_C^o$. The three different curves correspond to three different experimental systems, reported in Figure 5a of Heyman *et al.* (40) (top), which used S2 cells cotransfected with the expression plasmid A5C-hRXRα and the reporter plasmid ADH-CRBPII-LUC; Figure 4b of Levin *et al.* (41) (center), which used CV-1 cells cotransfected with the reporter CRBPII-RXRE-CAT construct and plasmid RXRα; and Figure 5b of Heyman *et al.* (40) (bottom), which used CV-1 cells cotransfected with the expression plasmid pRSh-RXRα and the reporter plasmid TK-CRBPII-LUC. The *NFI* was computed as $NFI_{R2} = \left(1 - (1+[s]/K_{lig})^{-4}\right)(1-P_t)$ (see Supplementary Methods) with equations (2) and (3). **(C)** Responses to atRA for systems with contiguous DNA binding sites for RXR. The three different dose-response curves correspond to the three systems of Figure 2B with the all-trans retinoic acid (atRA) as ligand of RXR instead of 9cRA. The highly variable dose-response curves are fully predicted without free parameters using the values of $\Delta G_C^o$ inferred in Figure 2B.

**Figure 3.** Local and global flexibility in the response landscapes. The normalized fold induction (*NFI*) for RXR from the model, computed as in Figure 2, is shown as a function of both the self-assembly modulator intensity (either 9cRA or atRA concentration) and the conformational free energy $\Delta G_C^o$. The figures on the bottom are density-plot projections of the corresponding *NFI* on the top, with dark and light gray corresponding to low and high values of the *NFI*, respectively. The red line corresponds to *NFI* =0.5 and shows the dependence of the half-maximum response concentration, or $EC_{50}$, with the conformational free energy. The $EC_{50}$ can be changed locally, at the single-promoter level, by changing the value of $\Delta G_C^o$, or globally, at a genome-wide scale by changing $K_{lig}$, the strength of the ligand binding to RXR. The 3D plots show the same experimental data (symbols) as in **(A)** Figure 2A, **(B)** Figure 2B, and **(C)** Figure 2C along with the dose-response curves (black lines) for the corresponding values of the conformational free energy. The values of the parameters used are $K_{td} = 4.4$ nM (7) and either $K_{lig} = 8$ nM for 9cRA (52) or $K_{lig} = 350$ nM for atRA (53).

**Figure 4.** Precision and flexibility of the transcriptional responses. The model for the whole-parameter space (see Supplementary Methods) reproduces the different dose-response curves regardless of 10-fold changes in total RXR concentration. The normalized fold induction from experimental data (symbols) and the model (lines) is shown for the same systems as in **(A)** Figure 2A, **(B)** Figure 2B, and **(C)** Figure 2C. The results agree with the experimental data over 10-fold changes of the total RXR concentration, given by $n_T = 4[n_4] + 2[n_2] + 2[n_2^*] + [n_1]$, which confirms the presence of a precise functional regime for typical molecular parameters. The experimental values used are $K_{lig} = 8$ nM for 9cRA (52) or $K_{lig} = 350$ nM for atRA (53); $K_{td} = 4.4$ nM and $K_{dm} = 155$ nM (7). The free energies of binding are obtained through their corresponding dissociation constants as $\Delta G_{s1}^o = RT \ln(K_{s1})$ and $\Delta G_{s2}^o = RT \ln(K_{s2})$ with $K_{s2} = K_{s1} = 8.8$ nM (54). $K_{dm}$, $K_{s1}$, and $K_{s2}$ are the dimer-monomer, RXR-site 1, and RXR-site 2 dissociation constants, respectively. The values of the conformational free energies were inferred in Figure 2. The 10-fold range of total RXR concentration was chosen for each specific system so that it best reproduces the experimental data in Figures 4A and 4B. The explicit value of $n_T$ is shown on the top of each graph (in nM units). The results of Figure 4C are predictions without free parameters.

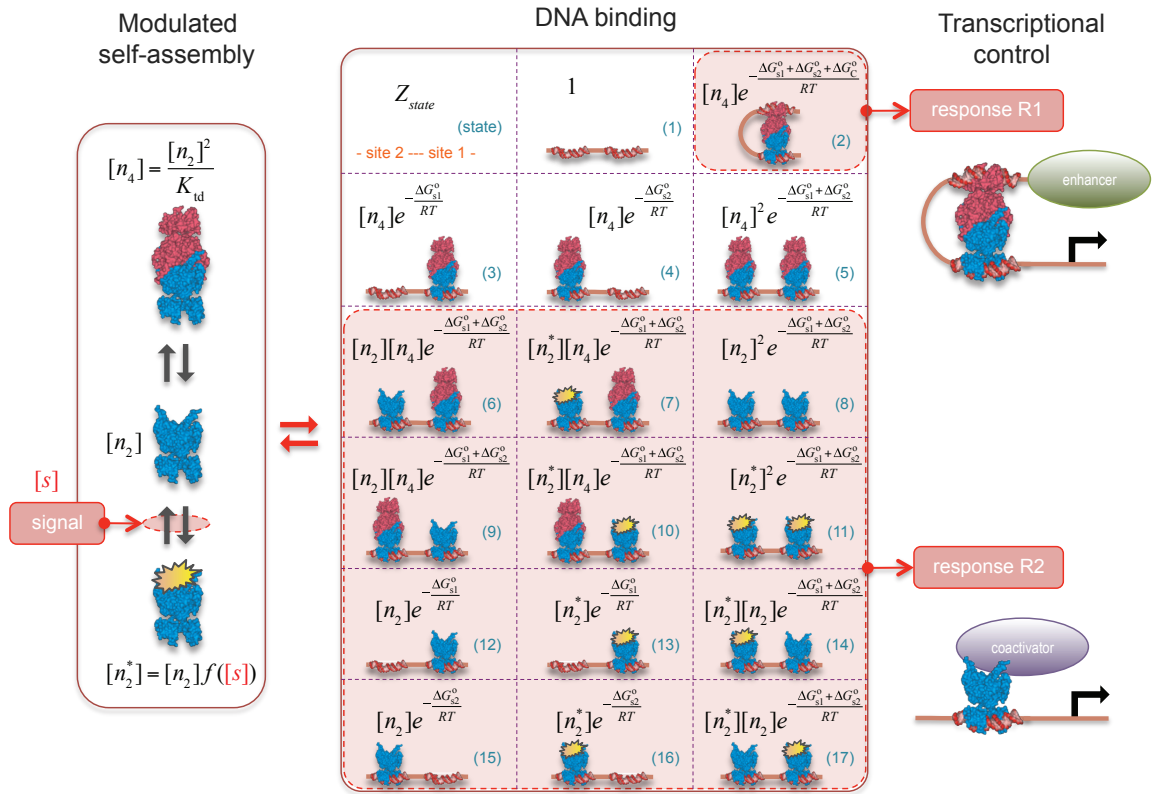

Figure 1

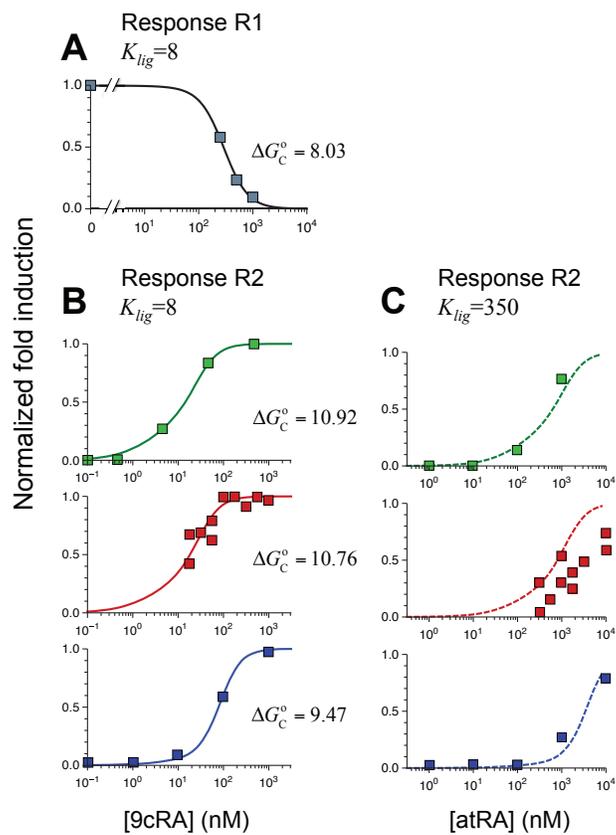

Figure 2

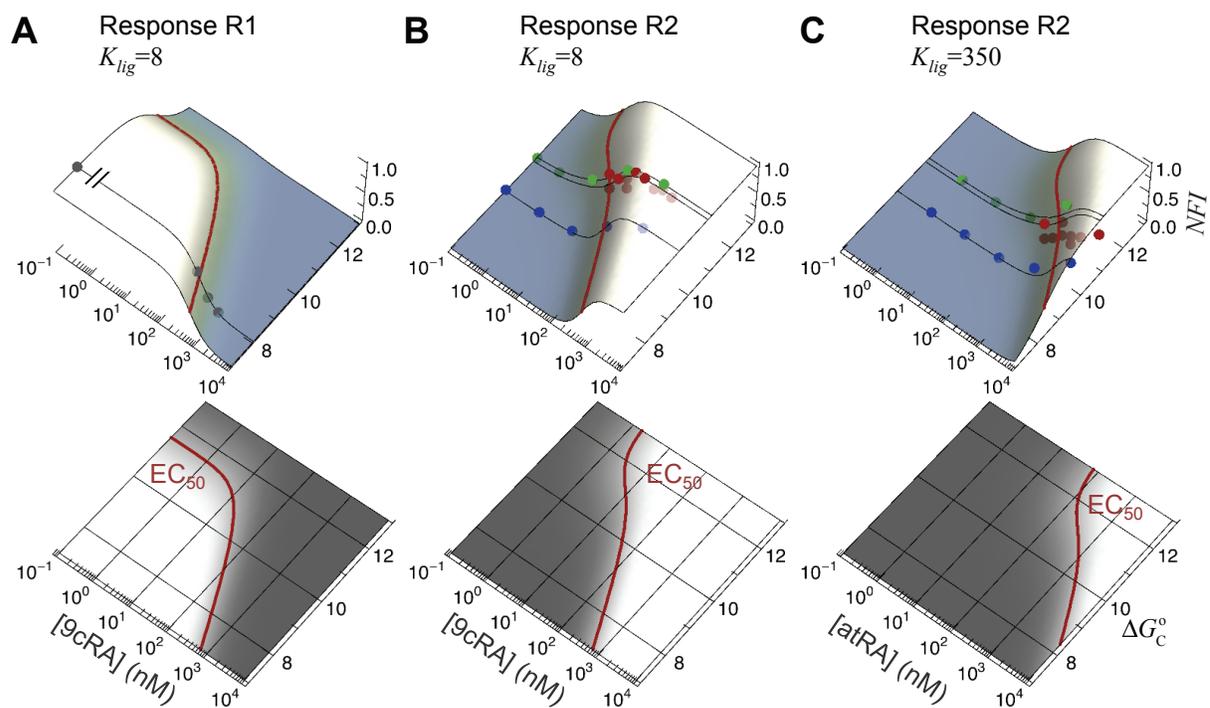

Figure 3

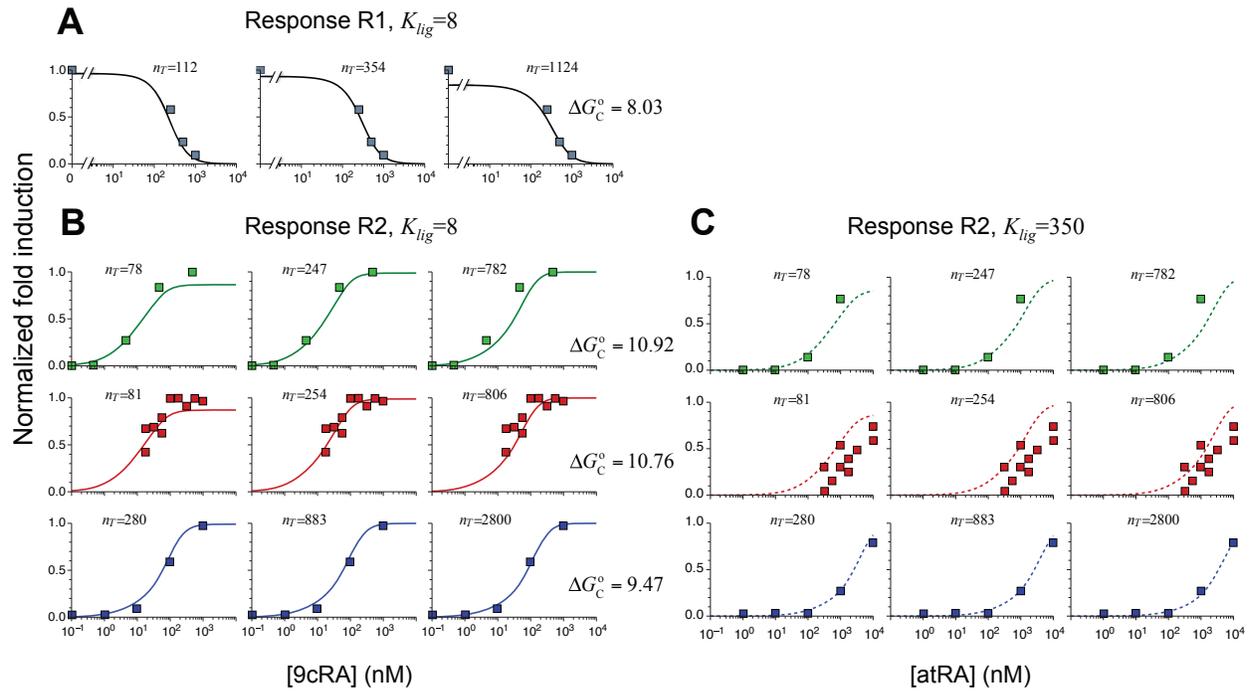

Figure 4

# SUPPLEMENTARY METHODS

## Normalized fold induction for response R1

In the case of an enhancer being brought to the promoter, the fold induction and maximum fold induction are $FI = \bar{\Gamma}_{R1} / \Gamma_{ref}$ and $FI_{max} = 1 + (\Gamma_t / \Gamma_{ref})$, respectively, which lead to a normalized fold induction given by

$$NFI_{R1} = P_t.$$

Therefore, in the case of response R1, the normalized fold induction is determined by just the probability of the conformation with the tetramer bound to the two DNA sites.

## Normalized fold induction for response R2

The case of a coactivator being recruited by an active RXR dimer is more involved because the self-assembly modulator, 9cRA, is also an agonist of RXR. The reason is that an RXR dimer has to bind at least one 9cRA molecule to recruit a coactivator. Therefore, the quantities $\Gamma_c$ (see text) are themselves effective transcription rates given by $\Gamma_c = \Gamma_{ref} P_{no\,ligand} + \Gamma_{act}(1 - P_{no\,ligand})$, where $P_{no\,ligand}$ is the probability that none of the dimers of the group of binding states $c$ has a ligand bound. It is given by $P_{no\,ligand} = 1/(1+[s]/K_{lig})^2$ for the groups of binding states with one dimer (od and do) and by $P_{no\,ligand} = 1/(1+[s]/K_{lig})^4$ for the group of binding states with the two dimers (dd). The resulting effective transcription rates for the groups of binding states with dimers —

$\Gamma_{od} = \Gamma_{ref} + (\Gamma_{act} - \Gamma_{ref})\left(1 - (1+[s]/K_{lig})^{-2}\right)$, $\Gamma_{do} = \Gamma_{ref} + (\Gamma_{act} - \Gamma_{ref})\left(1 - (1+[s]/K_{lig})^{-2}\right)$,

and $\Gamma_{dd} = \Gamma_{ref} + (\Gamma_{act} - \Gamma_{ref})\left(1 - (1+[s]/K_{lig})^{-4}\right)$ — depend explicitly on the self-assembly modulator concentration and take into account that binding of at least one ligand to RXR activates the dimeric form in addition to modulating the oligomerization state. In the case of response R2, therefore, we have $FI = \bar{\Gamma}_{R2}/\Gamma_{ref}$ and $FI_{max} = 1 + (\Gamma_{act}/\Gamma_{ref})$ and the resulting normalized fold induction is given by

$$NFI_{R2} = \left(1-(1+[s]/K_{lig})^{-2}\right)(P_{od} + P_{do}) + \left(1-(1+[s]/K_{lig})^{-4}\right)P_{dd}$$

for the whole-parameter space, and by

$$NFI_{R2} \approx \left(1-(1+[s]/K_{lig})^{-4}\right)(1-P_t)$$

in the functional regime.

## Modulator function for RXR

The self-assembly modulator of RXR is the hormone 9-*cis*-retinoic acid (9cRA), a derivative of Vitamin A, which binds each RXR molecule independently of its oligomerization state and prevents dimers with their two subunits occupied from tetramerazing. Therefore, there are several types of tetramerizing dimers. We use the notation $[n_{0,0}]$ for the dimer concentration with no hormone bound; $[n_{0,1}]$ and $[n_{1,0}]$ for those with just one 9cRA molecule bound; and $[n_{1,1}]$ for those with two 9cRA molecules bound. Binding of 9cRA to RXR follows the usual mass action law: $[n_{1,0}] = [n_{0,0}][s]/K_{lig}$, $[n_{0,1}] = [n_{0,0}][s]/K_{lig}$, and $[n_{1,1}] = [n_{0,0}][s]^2/K_{lig}^2$, where $K_{lig}$ is the ligand-RXR dissociation constant. The concentrations of tetramerazing and non-tetramerazing dimers are therefore related to each other by

$$[n_2] = [n_{1,0}] + [n_{0,1}] + [n_{0,0}] = [n_{0,0}](2[s]/K_{lig} + 1)$$

$$[n_2^*] = [n_{1,1}] = [n_{0,0}][s]^2 / K_{lig}^2$$

from which we obtain the explicit form of the modulator function:

$$f([s]) = \frac{[n_2^*]}{[n_2]} = \frac{[s]^2}{K_{lig}^2 + 2K_{lig}[s]}.$$

This expression explicitly indicates how the ligand controls the relative concentrations of the different oligomerization states that shape the transcriptional response.

**Computational approach for RXR in the whole-parameter space**

The whole-parameter space needs to consider explicitly the total nuclear RXR concentration, $n_T = 4[n_4] + 2[n_2] + 2[n_2^*] + [n_1]$, which includes the contributions from monomer concentration $[n_1]$ in addition to those from the tetramer, dimer, and non-tetramerizing dimer. Dimerization is described by $[n_1]^2 / ([n_2] + [n_2^*]) = K_{dm}$, where $K_{dm}$ is dimer-monomer dissociation constant. For given values of the parameters, the concentrations of the four oligomerization states are obtained by solving numerically the equations for the total nuclear RXR concentration, dimerization, tetramerization, and modulator function for each ligand concentration. The resulting oligomeric concentrations are used to obtain the corresponding probabilities (Table 2), which in turn are substituted in the expressions $NFI_{R1} = P_t$ and $NFI_{R2} = \left(1 - (1+[s]/K_{lig})^{-2}\right)(P_{od} + P_{do}) + \left(1 - (1+[s]/K_{lig})^{-4}\right)P_{dd}$ to obtain the normalized fold induction for responses R1 and R2, respectively.